\documentclass{mn2e}
\usepackage{graphicx}
\newcommand{\Fs}{\,^*\! F}

\newcommand{\bJ}{\bmath{J}}

\newcommand{\br}{\bmath{r}}
\newcommand{\bv}{\bmath{v}}

\newcommand{\bg}{\bmath{g}}

\newcommand{\bP}{\bmath{P}}
\newcommand{\bB}{\bmath{B}}
\newcommand{\bE}{\bmath{E}}

\newcommand{\bS}{\bmath{S}}

\newcommand{\text}[1]{\quad\mbox{#1}\quad}
\newcommand{\spr}[2]{\bmath{#1} \!\cdot\! \bmath{#2}}
\newcommand{\vpr}[2]{\bmath{#1} \!\times\! \bmath{#2}}
\newcommand{\vdiv}[1]{\spr{\nabla}{#1}}
\newcommand{\vgrad}[1]{\nabla{#1}}
\newcommand{\vcurl}[1]{\vpr{\nabla}{#1}}

\newcommand{\pder}[2]{\frac{\partial #1}{\partial #2}}
\newcommand{\Pd}[1]{\partial_{#1}}

\title{Multi-dimensional Numerical Scheme for Resistive Relativistic MHD}
\author[S.~S. Komissarov]{Serguei S.~Komissarov
\thanks{E-Mail:~serguei@maths.leeds.ac.uk}\\
Department of Applied Mathematics, The University of Leeds,
Leeds, LS2 9GT}
\begin{document}
\date{Received/Accepted}
\maketitle

\begin{abstract}
The paper describes a new upwind conservative numerical scheme 
for special relativistic resistive magnetohydrodynamics with scalar resistivity. 
The magnetic field is kept approximately divergence free and the divergence of the 
electric field consistent with the electric charge distribution via the method of 
Generalized Lagrange Multiplier. The hyperbolic fluxes are computed using the HLL 
prescription and the source terms are accounted via the time-splitting technique. 
The results of test simulations show that the scheme can handle equally well both 
resistive current sheets and shock waves and thus can be a useful tool for studying 
phenomena of relativistic astrophysics that involve both colliding supersonic flows 
and magnetic reconnection.

\end{abstract}
                                                                                          
                                                                                          
\section{Introduction}
\label{introduction}

In many phenomena of relativistic astrophysics, such as AGN, GRBs, quasars, radio galaxies, 
micro-quasars, pulsars and magnetars, compact X-ray binaries etc., the magnetic field is a key 
dynamic component. On one hand, the magnetic field drives, accelerates and partially collimates 
relativistic outflows from astrophysical black holes, neutron stars, and their accretion disks. 
On the other hand, magnetic reconnection and dissipation is responsible for bright thermal and 
non-thermal emission from these flows. Recent years have seen a remarkable progress in numerical 
methods for ideal relativistic magnetohydrodynamics 
\cite{Kom99,KSK99,Kom01,KKU,GMT03,Due05,KSK99,Ann06,SS05,AHLN06,Del03,Ant06,NHM06,Miz06,MB06,M06,N06,GR07,Del07} 
and many interesting and important simulations have been carried out already. Quite often the numerical  
solutions exhibited violent magnetic reconnection.  Although it is indeed very likely to occur in the 
considered astrophysical phenomena   
as the result of non-vanishing physical resistivity of plasma, both collisional and collisionless, 
the reconnection observed in the simulations is of purely numerical origin. It is driven by artificial 
resistivity arising due to truncation errors and hence fully depending on fine details of numerical 
schemes and resolution. A code for resistive RMHD would allow to control magnetic reconnection 
according to the incorporated 
physical models of resistivity. Moreover, the relativistic magnetic reconnection by itself is a 
sufficiently rich and important physical process to warrant the effort of developing such a code. 
The only numerical study of relativistic magnetic reconnection  so far was carried out by 
Watanabe \& Yokoyama\shortcite{WY06}. However, their paper gives no details of their numerical scheme 
and test simulations and therefore it is not clear as to how accurate their results are and how robust 
their numerical method is.  

Since many relevant astrophysical phenomena involve shock waves, including the 
fast magnetic reconnection of Petcheck type \cite{Lyub05},    
a useful code should handle well not only current sheets and filaments but also shock waves. 
It is well known that codes that do not preserve the magnetic field divergence free can become 
unstable and crash in the cases with large spacial gradients. Thus this issue must be 
addressed too. Moreover, in the relativistic limit the spacial charge density and the advective 
current can become significant and thus the electric charge conservation has to be enforced.    
In this paper we describe the results of our efforts to construct a code that satisfies these
criteria. The equations of resistive RMHD are described in Section~\ref{equations}. 
The equations of 
the so-called augmented system of resistive RMHD, that are designed to handle to enforce the 
divergence free condition for magnetic field and the electric charge conservation are presented 
in Section~\ref{au-equations}. The relativistic Ohm law is explained in Section~\ref{Ohm law}. 
Section~\ref{num-method} gives the details of numerical integration. 
The test simulations are described in Section~\ref{tests} and our conclusions 
are summarized in Section~\ref{conclusions}.

\section{Basic equations}
\label{equations}
                
The covariant Maxwell equations are \cite{JAC}
                                                                                     
\begin{equation}
  \nabla_\beta  \Fs^{\alpha \beta} = 0,
\label{Maxw1}
\end{equation}
                                                                                     
\begin{equation}
  \nabla_\beta  F^{\alpha \beta} =  I^\alpha,
\label{Maxw2}
\end{equation}
                                                                                     
\noindent
where $F^{\alpha\beta}$ is the Maxwell tensor of the electromagnetic
field, $\Fs^{\alpha\beta}$ is the Faraday tensor, and
$I^\alpha$ is the 4-vector of electric current.

In highly ionized plasma, including pair plasma, 
the electric and magnetic susceptibilities are essentially zero and one has 
                                                                                     
\begin{equation}
\Fs^{\alpha  \beta} = \frac{1}{2} e^{\alpha \beta \mu \nu} F_{\mu \nu}
\label{dual_F}
\end{equation}
                                                                                     
\begin{equation}
F^{\alpha  \beta} = -\frac{1}{2} e^{\alpha \beta \mu \nu} \Fs_{\mu \nu},
\label{F}
\end{equation}
where
                                                                                     
\begin{equation}
e_{\alpha \beta \mu \nu} = \sqrt{-g}\,\epsilon_{\alpha \beta \mu \nu},
\label{LCt}
\end{equation}
                                                                                     
\noindent
is the Levi-Civita alternating tensor of space-time
and  $\epsilon_{\alpha \beta \mu \nu}$ is the four-dimensional Levi-Civita
symbol.

In the coordinate basis of global inertial frame of special relativity these equations 
split into the familiar set 

\begin{equation}
   \vdiv{B}=0,
\label{GaussB}
\end{equation}
                                                                                     
\begin{equation}
   \Pd{t}\bB + \vcurl{E} = 0,
\label{Faraday}
\end{equation}

\begin{equation}
\label{GaussE}
   \vdiv{E}=q,
\end{equation}
                                                                                     
\begin{equation}
   -\Pd{t}\bE + \vcurl{B} = \bJ,
\label{Ampere}
\end{equation}

\noindent
where
\begin{equation}
     E^i= F^{ti} =\frac{1}{2} e^{ijk} \Fs_{jk},
\label{E1}
\end{equation}
\begin{equation}
     B^i= \Fs^{it} = \frac{1}{2}e^{ijk} F_{jk},
\label{B1}
\end{equation}
\begin{equation}
    q= I^t, \quad J^k=I^k.
\label{rhoJ}
\end{equation}
are the electric field, the magnetic field, the electric charge density, and 
the electric current density respectively as measured by the inertial observer
($e_{ijk} = e_{0ijk}$ is the Levi-Civita tensor of space.).
These equations are consistent with the electric charge conservation

\begin{equation}
    \Pd{t}q + \vdiv{J}= 0.
\label{ECC}
\end{equation}

In magnetohydrodynamics Maxwell's equations are supplemented with the equations
of motion of matter and the continuity equation. 
In the covariant form the equations of motion can be written as

\begin{equation}
     \nabla_\nu T^{\mu\nu} =0 
\label{EMc}
\end{equation}
where the total stress-energy momentum tensor,

\begin{equation}
     T^{\mu\nu} = T_{(m)}^{\mu\nu} + T_{(e)}^{\mu\nu},
\label{tsemt}
\end{equation}
is the sum of the stress-energy momentum tensor of the electromagnetic field

\begin{equation}
   T_{(e)}^{\mu\nu} = F^{\mu\gamma} F^\nu_{\ \gamma} -
   \frac{1}{4}(F^{\alpha\beta}F_{\alpha\beta})g^{\mu\nu},
\label{semtef}
\end{equation}
and the stress-energy momentum tensor of matter 

\begin{equation}
   T_{(m)}^{\mu\nu} = wu^\mu u^\nu + p g^{\mu\nu}.
\label{semtm}
\end{equation}
Here where $p$ is the thermodynamic pressure, $w(p,\rho)$ is the relativistic enthalpy per
unit volume as measured in the rest frame of fluid ($w$ includes the rest mass-energy 
density of matter $\rho$), and $u^\nu$ is the fluid 4-velocity. In the global 
inertial frame with time-independent coordinate grid 
eq.\ref{EMc} splits into the energy and momentum conservation laws                                                                                     
\begin{equation}
   \Pd{t}{e}+\vdiv{S}=0,
\label{ecl}
\end{equation}
\begin{equation}
   \Pd{t}{\bP}+\vdiv{\Pi}=0,
\label{mcl}
\end{equation}
where 
\begin{equation}
   e=\frac{1}{2}(E^2+B^2)+w\gamma^2-p
\end{equation}
is the energy density, 
\begin{equation}
   \bS=\vpr{E}{B}+w\gamma^2 \bv,
\end{equation}
is the energy flux density,
\begin{equation}
   \bP=\vpr{E}{B}+w\gamma^2 \bv
\end{equation}
is the momentum density, and 
\begin{equation}
   \bmath{\Pi}=-\bE\bE-\bB\bB + w\gamma^2 \bv\bv +\left(\frac{1}{2}(E^2+B^2)+p\right) \bg 
\end{equation}
is the stress tensor. Here $\gamma$ is the Lorentz factor, $\bv$ is the velocity 
as measured by the inertial observer, and $\bg$ is the metric tensor of space. 

The covariant continuity equation is 

\begin{equation}
   \nabla_\nu {\rho u^\nu}=0,
\end{equation}
where $\rho$ is the rest mass density as measured in the rest frame of fluid. 
In the inertial frame this reads 
\begin{equation}
   \Pd{t}{\rho\gamma} + \bmath{\nabla}\cdot (\rho\gamma \bv)=0.
\label{ce}
\end{equation}

Equations 
(\ref{GaussB},\ref{Faraday},\ref{GaussE},\ref{Ampere},\ref{ECC},\ref{ecl},\ref{mcl},\ref{ce}) 
constitute the 3+1 PDE system of 
relativistic magnetohydrodynamics in special relativity. Once supplemented 
with equations of state, that relate various thermodynamic parameters of matter, and 
with the Ohm law, that couples matter and the electromagnetic field, this system closes.

\section{Augmented system}
\label{au-equations}

As well known, the divergence free condition (\ref{GaussB}) for the magnetic field can 
be treated as a constraint on the initial solution of Cauchy problem because the Faraday 
equation (\ref{Faraday}) will then ensure that the magnetic field remains divergence 
free at any time. The equation of electric charge conservation is also not 
independent and follows from the Ampere equation (\ref{Ampere}) and the 
Gauss law (\ref{GaussE}). These properties of the differential equations 
are not preserved by many numerical schemes. Indeed, the most straightforward way of 
constructing a self-consistent finite difference counterpart for a differential system
like electrodynamics is to ignore all constraints (non-evolution equations) and to 
leave out all the supplementary laws like the electric charge conservation (otherwise 
the system of finite difference equations becomes over-determined).
However, it has been discovered that this lack of consistency may lead to strong corruption 
of numerical solutions in regions with large truncation errors, like 
strong discontinuities, and even cause code crash. In ideal MHD the divergence 
free condition has been found particularly important. A number of techniques 
has been proposed to combat the problem. Here we adopt the so-called 
Generalized Lagrange Multiplier method developed by Munz et al.\shortcite{Munz99}.         
The main idea is to create a new, augmented system of differential equations, that will 
include only evolution equations and will have the same solutions of the Cauchy problem 
as the original system provided the initial solution satisfies the differential 
constraints of the original system. If, however, the initial solution does not 
satisfy the constraints then the deviations should decay or at least 
move away as relatively high speed waves. This will ensure that the deviations caused by 
truncation errors of a numerical method for the augmented system remain small.       

To deal with the divergence free constraint we modify eqs.(\ref{GaussB},\ref{Faraday})
so that they become 
                                                                                                
\begin{equation}
   \Pd{t}\Phi + \vdiv{B}=-\kappa\Phi,
\label{aGaussB}
\end{equation}
                                                                                                
\begin{equation}
   \Pd{t}\bB + \vcurl{E} + \vgrad{\Phi} = 0,
\label{aFaraday}
\end{equation}
where $\Phi$ is a new dynamic variable (pseudo-potential). From these equations 
it follows that $\Phi$ satisfies the telegraph equation 

\begin{equation}
   -\Pd{t}^2\Phi -\kappa\Pd{t}\Phi + \nabla^2 \Phi = 0. 
\label{tel-phi}
\end{equation}
Thus, $\Phi$ is transported by hyperbolic waves propagating with the speed of light 
and decays if $\kappa>0$. For positive $\kappa$ the natural evolution of $\Phi$ is 
toward $\Phi(\br,t)=0$ 
(unless prevented by boundary conditions) and eq.(\ref{aGaussB}) shows that this final 
state implies divergence free magnetic field. In fact, it is easy to see that the 
divergence of magnetic field also satisfies the same telegraph equation 

\begin{equation}
   -\Pd{t}^2(\vdiv{B}) -\kappa\Pd{t}(\vdiv{B}) + \nabla^2 (\vdiv{B}) = 0,
\label{tel-divb}
\end{equation}
and thus evolves in the same fashion.     

To deal with the Gauss law we modify eqs.(\ref{GaussE},\ref{Ampere}) so they read 
                                                                                                
\begin{equation}
   \Pd{t}\Psi + \vdiv{E}=q-\kappa\Psi,
\label{aGaussE}
\end{equation}
                                                                                                
\begin{equation}
   -\Pd{t}\bE + \vcurl{B} -\vgrad{\Psi} = \bJ,
\label{aAmpere}
\end{equation}
where $\Psi$ is another new dynamic variable. From these two equations and 
the electric charge conservation it follows that the evolution of $\Psi$ is  again
described by the telegraph equation 
                                                                                                
\begin{equation}
   -\Pd{t}^2\Psi -\kappa\Pd{t}\Psi + \nabla^2 \Psi = 0.
\label{tel-psi}
\end{equation}
(Although in principle one could use 
different constants $\kappa$ for $\Phi$ and $\Psi$ this brings no benefit.) 
Thus, $\Psi$, naturally evolves in the same fashion as $\Phi$, ensuring that 
the electrodynamic solution is kept consistent with the Gauss law. Similarly,
one finds that 

\begin{equation}
   -\Pd{t}^2(\vdiv{E}-q) -\kappa\Pd{t}(\vdiv{E}-q) + \nabla^2 (\vdiv{E}-q) = 0.
\label{tel-dive}
\end{equation}

Summarizing, the augmented system of relativistic MHD is  
                                                                                                
\begin{equation}
   \Pd{t}\Phi + \vdiv{B}=-\kappa\Phi,
\label{eeq1}
\end{equation}

\begin{equation}
   \Pd{t}\bB + \vcurl{E} + \vgrad{\Phi} = 0,
\end{equation}

\begin{equation}
   \Pd{t}\Psi + \vdiv{E}=q-\kappa\Psi,
\end{equation}

\begin{equation}
   -\Pd{t}\bE + \vcurl{B} -\vgrad{\Psi} = \bJ,
\end{equation}
 
\begin{equation}
    \Pd{t}q + \vdiv{J}= 0.
\end{equation}

\begin{equation}
   \Pd{t}\rho \gamma + \nabla\cdot\rho\gamma\bv =0,
\end{equation}

\begin{equation}
   \Pd{t}{e}+\vdiv{S}=0,
\end{equation}

\begin{equation}
   \Pd{t}{\bP}+\vdiv{\Pi}=0,
\label{eeq2}
\end{equation}

where
\begin{equation}
   e=\frac{1}{2}(E^2+B^2)+w\gamma^2-p
\end{equation}
\begin{equation}
   \bS=\vpr{E}{B}+w\gamma^2 \bv,
\end{equation}
\begin{equation}
   \bP=\vpr{E}{B}+w\gamma^2 \bv
\end{equation}
\begin{equation}
   \bmath{\Pi}=-\bE\bE-\bB\bB + w\gamma^2 \bv\bv +\left(\frac{1}{2}(E^2+B^2)+p\right) \bg
\end{equation}
Every differential equation of the system is an evolution equation and a conservation 
law (with or without a source term), and there is wealth of numerical methods 
for such systems. For example one could use Godunov's upwind scheme \cite{God} or one of 
its numerous higher order ``children''. 
This simplicity of numerical implementation is the main advantage of the 
method of Generalized Lagrange Multiplier.

\section{Ohm's law}
\label{Ohm law}

In this paper we consider only the simplest case of scalar resistivity. In strong 
magnetic field the resistivity (conductivity) becomes anisotropic and the tensor description
becomes more appropriate. We will consider this case in future. 
   
The covariant form of scalar Ohm's law is 

\begin{equation}
  I_\nu = \sigma F_{\nu\mu}u^\mu + q_0u_\nu,
\label{cOL}
\end{equation}
where $\sigma=1/\eta$ is the conductivity, $\eta$ is the resistivity, and 
$q_0 = -I_\nu u^\nu$ is the electric charge density as measured in the fluid frame \cite{BF93,LU03}.
In the general inertial frame this reads 

\begin{equation}
  \bJ = \sigma\gamma\left[ \bE + \vpr{v}{B} -(\spr{E}{v})\bv \right] + q \bv,
\label{OHML}
\end{equation}
whereas in the fluid frame one has the usual Ohm law
\begin{equation}
  \bJ = \sigma \bE. 
\end{equation}
In the limit of infinite conductivity ($\sigma\to\infty$) eq.(\ref{OHML}) reduces to 
$$
\bE + \vpr{v}{B} -(\spr{E}{v})\bv =0.  
$$
Splitting this equation into the components that are normal and parallel to 
the velocity vector one obtains  
$$
  \bE_\perp+\vpr{v}{B}=0
$$
and 
$$
 \bE_\parallel -(\spr{E}{v})\bv=0.
$$
These show that $E_\parallel=0$ and thus one has the usual result 
\begin{equation}
\bE = -\vpr{v}{B},
\end{equation}
the purely inductive electric field.

Now consider the reduced form of Ampere's law 
\begin{equation}
   -\Pd{t}\bE = \bJ,
\label{Ampere-1}
\end{equation}
which is of interest for numerical schemes using time-splitting 
technique. When splitted into components normal and parallel to
the velocity vector this equation reads 
\begin{equation}
   \Pd{t}\bE_\parallel + \sigma\gamma[\bE_\parallel -(\spr{E}{v})\bv]=0,
\end{equation}
\begin{equation}
   \Pd{t}\bE_\perp + \sigma\gamma[\bE_\perp + \vpr{v}{B}]=0.
\end{equation}
The solutions of initial value problem for these linear equations are 
\begin{equation}
   \bE_\parallel = \bE_\parallel^0 \exp{\left(-\frac{\sigma}{\gamma}t\right)},
\label{sol-para}
\end{equation}
and
\begin{equation}
   \bE_\perp = \bE_\perp^* +(\bE_\perp^0-\bE_\perp^*)\exp{(-\sigma\gamma t)},
\label{sol-perp}
\end{equation}
where $\bE_\perp^*=-\vpr{v}{B}$ and suffix $0$ denotes the initial values. 
One can see that for relativistic flows the normal component of 
electric field approaches the inductive value $\bE_\perp^*$ faster 
than the parallel component approaches zero.

\section{Numerical method}
\label{num-method}

The evolution equations (\ref{eeq1}-\ref{eeq2}) can be written as
conservation laws. In Cartesian coordinates, and this is the only type of coordinates 
we use in the paper, the system can be written as a single phase vector equation 
                                                                                          
\begin{equation}
  \pder{{\cal Q}({\cal P})}{t} + \pder{{\cal F}^{m}({\cal P})}{x^m}  = {\cal S}({\cal P}),
\label{CONS}
\end{equation}
where

$$
  {\cal Q} = \left(
   \begin{array}{c} 
       \Phi\\ B^i\\ \Psi\\ E^i\\ q \\ \rho\gamma\\ e\\ P^i 
   \end{array} 
   \right), \quad
  {\cal P} = \left(
   \begin{array}{c} 
       \Phi\\ B^i\\ \Psi\\ E^i\\ q\\ \rho\\ p\\ u^i 
   \end{array} 
   \right), \quad
 {\cal S} = \left(
   \begin{array}{c}
           -\kappa \Psi\\0^i\\ q-\kappa \Psi\\ -J^i\\0\\0\\0\\0^i
   \end{array}
   \right)
$$
are the vectors of conserved quantities, primitive quantities, and sources 
respectively and 

$$                                                                                         
 {\cal F}^{m} = \left( 
   \begin{array}{c}
       B^m\\ e^{imk}E_k+\Phi g^{im}\\  E^j\\  - e^{imk}B_k+\Psi g^{im}\\ 
                         J^m\\ \rho u^m\\ S^m\\\Pi^{im} 
   \end{array}
   \right)
$$
is the vector of corresponding hyperbolic fluxes.
Here $u^i=\gamma v^i$ are the spatial components of 4-velocity, $g^{ij}$ are the
components of the metric tensor of space (given by Kronecker's delta $\delta^{ij}$),
$e^{ijk}$ is the Levi-Civita alternating tensor of space.  

We have found useful to split the source term into two parts
$$
  {\cal S}_a({\cal P}) = \left(
   \begin{array}{c}
          0\\ 0^i\\ q\\ -qv^i\\0\\0\\0\\0^i
   \end{array}
   \right) \text{and}
  {\cal S}_b({\cal P}) = \left(
   \begin{array}{c}
        -\kappa \Phi\\0^i\\-\kappa \Psi\\ -J_c^i\\0\\0\\0\\0^i
   \end{array}
   \right),
\label{Sa}
$$
where  
$$
  \bJ_c = \sigma\gamma\left[ \bE + \vpr{v}{B} -(\spr{E}{v})\bv \right]
$$
is the conductivity current.  The source term ${\cal S}_b$ is potentially 
stiff (in the case of low resistivity) and is treated via the time-step splitting
technique by Strang \shortcite{Str}. That is first the solution is advanced 
via integration of equation 
                                                                                                 
\begin{equation}
  \pder{{\cal Q}({\cal P})}{t} = {\cal S}_b({\cal P}),
\label{step1}
\end{equation}
over the half time-step, $\Delta t/2$. 
Then the solution is advanced via second-order accurate numerical integration
of equation 
                                                                                                 
\begin{equation}
  \pder{{\cal Q}({\cal P})}{t} + \pder{{\cal F}^{m}({\cal P})}{x^m}  = 
                             {\cal S}_a({\cal P}),
\label{step2}
\end{equation}
over the full time-step. Finally, the solution is advanced
via integration of equation (\ref{step1}) over the half time-step once more.  
Thanks to the fact that all equations in (\ref{step1})
are linear the integration is carried out using analytic solutions, in particular 
solutions (\ref{sol-para},\ref{sol-perp}) are utilized at this stage. This removes 
stability constraints of the time step otherwise imposed by ${\cal S}_b$.    
In principle all source terms could be passed to eq.(\ref{step1}) but this somehow 
results in reduction of accuracy.

\begin{figure*}
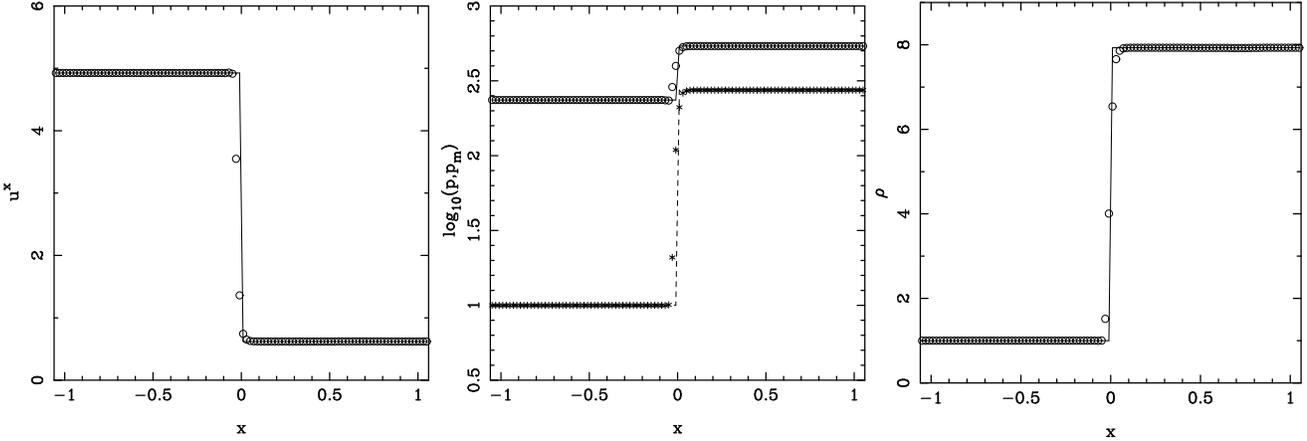

\includegraphics[angle=-90,width=57mm]{figures/sfs-u.eps}
\includegraphics[angle=-90,width=57mm]{figures/sfs-p.eps}
\includegraphics[angle=-90,width=57mm]{figures/sfs-d.eps}
\caption{ Stationary fast shock. {\it Left panel:} $u^x=\gamma v^x$; 
{\it Middle panel:} gas pressure (dashed line and stars) and magnetic pressure 
(solid line and circles); {\it Right panel} rest mass density, $\rho$.
}
\label{sfs}
\end{figure*}
\begin{figure*}
\includegraphics[angle=-90,width=57mm]{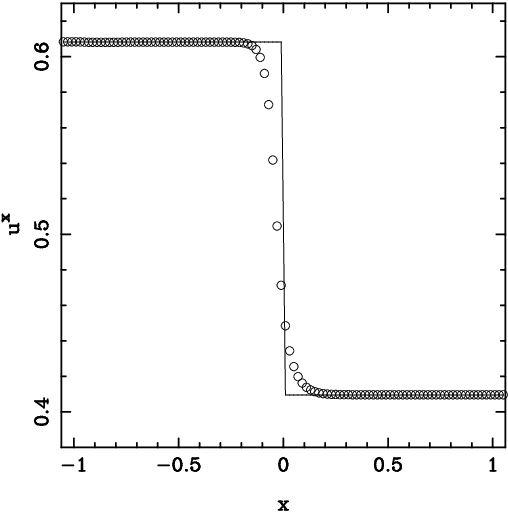}
\includegraphics[angle=-90,width=57mm]{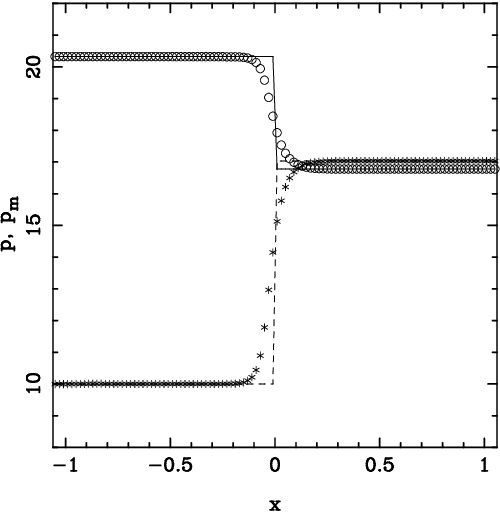}
\includegraphics[angle=-90,width=57mm]{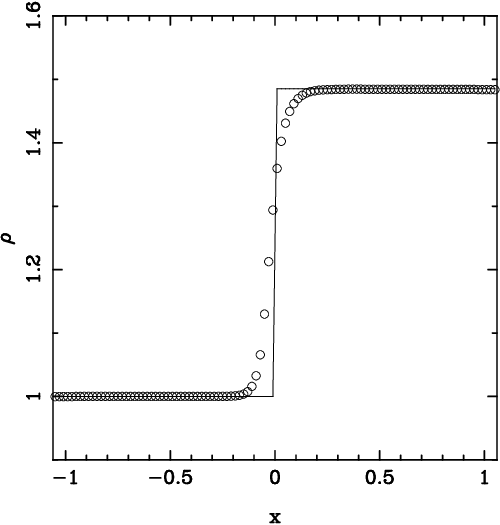}
\caption{ Stationary slow shock. {\it Left panel:} $u^x=\gamma v^x$;
{\it Middle panel:} gas pressure (dashed line and stars) and magnetic pressure
(solid line and circles); {\it Right panel} rest mass density, $\rho$.
}
\label{sss}
\end{figure*}
  
Equation~(\ref{step2}) is integrated explicitly 

$$
  {\cal Q}_{n+1} =  {\cal Q}_{n}  + 
        \Delta t\sum\limits_{m=1}^{N_d} \frac{{\cal F}_{m-\frac{1}{2},n+\frac{1}{2}}  
          - {\cal F}_{m+\frac{1}{2},n+\frac{1}{2}}}{\Delta x^m}
$$
\begin{equation}
     \qquad\qquad\qquad\qquad\qquad\qquad\qquad\qquad\qquad 
     + \Delta t\, {\cal S}_{a,n+\frac{1}{2}}. 
\label{step2-int}
\end{equation}
Here ${\cal Q}_{n}$ is the conserved quantity of a cell at $t=t_n$, 
${\cal Q}_{n+1}$ is the conserved quantity of this cell at $t=t_n+\Delta t$,
${\cal S}_{a,n+\frac{1}{2}}$ is the source term of the cell at $t=t_n+\Delta t/2$, 
${\cal F}_{m+\frac{1}{2},n+\frac{1}{2}}$ is the flux though the right 
interface and ${\cal F}_{m-\frac{1}{2},n+\frac{1}{2}}$ is the flux though 
the left interface of the cell normal to the direction of $x^m$ at
time $t=t_n+\Delta t/2$. $\Delta x^m$ is the cell size in this direction 
and $N_d$ is the number of spatial dimensions. 

To determine the sources and fluxes at half time-step the solution is temporary 
advances via 
$$
  {\cal Q}_{n+\frac{1}{2}} =  {\cal Q}_{n}  +
        \frac{\Delta t}{2}\sum\limits_{m=1}^{N_d} \frac{{\cal F}_{m-\frac{1}{2},n}
          - {\cal F}_{m+\frac{1}{2},n}}{\Delta x^m}
$$
\begin{equation}
     \qquad\qquad\qquad\qquad\qquad\qquad\qquad\qquad\qquad
     + \frac{\Delta t}{2} {\cal S}_{a,n}.
\label{step2a-int}
\end{equation}
The interface fluxes ${\cal F}_{m+\frac{1}{2},n}$ are computed using the
HLL-prescription \cite{HLL}:
\begin{equation}
   {\cal F}_{m+\frac{1}{2},n} = 
      \frac{{\cal F}^R_{m+\frac{1}{2},n} + {\cal F}^L_{m+\frac{1}{2},n}}{2} - 
      \frac{{\cal Q}^R_{m+\frac{1}{2},n} - {\cal Q}^L_{m+\frac{1}{2},n}}{2},   
\label{HLL}
\end{equation}
where indexes $L$ and $R$ refer to the states respectively to the left and to the 
right of the interface (which can be considered as the location of discontinuity in 
the solution at $t=t_n$). Note the simplification of the general HLL prescription 
due to the fact that the maximum characteristic speed of the system in each direction 
equals exactly to the speed of light (unity in our dimensionless equations).   
For the auxiliary half time-step these left and right states are found via the 
piece-wise constant reconstruction of numerical solution in each spatial direction 

\begin{equation}
   {\cal P}_{n} = {\cal P}_{n}^c \text{for} 
   x_c^m-\frac{\Delta x^m}{2} < x^m < x_c^m+\frac{\Delta x^m}{2},  
\label{pcr}
\end{equation}
where $x_c^m$ is the coordinate of the cell center and ${\cal P}_{n}$ is the phase 
state vector of the cell.    
 
The auxiliary solution is then used for another, now quadratic reconstruction of 
numerical solution within each cell 

$$
   {\cal P}_{n+\frac{1}{2}} = {\cal P}_{n+\frac{1}{2}}^c + 
                  a_1(x^m-x^m_c) +\frac{a_2}{2}(x^m-x^m_c)^2 
$$
\begin{equation}
\text{for}  x_c^m-\frac{\Delta x^m}{2} < x^m < x_c^m+\frac{\Delta x^m}{2}.
\label{pcr-1}
\end{equation}
Obviously, $a_1$ and $a_2$ are the first and the second order derivatives of the 
reconstructed solution and these are to be found from the numerical solution using one of many
existing non-linear limiters (needed to avoid spurious oscillations). In this particular paper 
$a_1$ is found using the same limiter as in our ideal MHD code \cite{Kom99}  

\begin{equation}
  a_1 = \mbox{av}\left(
        {\cal P}'_L,
        {\cal P}'_R
      \right)
\end{equation}
where 

$$
   {\cal P}'_L= 
        \frac{{\cal P}_{i}-{\cal P}_{i-1}}{\Delta x^m},
   \quad
   {\cal P}'_R= 
        \frac{{\cal P}_{i+1}-{\cal P}_{i}}{\Delta x^m},
$$
are the left and right numerical approximations of the first derivative 
($i$ is the cell index along the direction of $x^m$), and

\begin{equation} 
  \mbox{av}(a,b)=\left\{ 
  \begin{array}{ccl} 
     0 &\mbox{if}& ab<0 \text{or } a^2 + b^2= 0,\\
     \frac{a^2 b + ab^2}{a^2 + b^2} &\mbox{if}& ab\ge 0 \text{and} a^2 + b^2\ne 0
  \end{array} \right.
\end{equation} 
To find $a_2$ we use a similar procedure. First we compute the left, center, and right numerical 
approximations for the second derivative, ${\cal P}''_L$, ${\cal P}''_C$, and ${\cal P}''_R$, 
and then we feed them to the minmod function with three arguments 

\begin{equation}
  a_2=\mbox{minmod}({\cal P}''_L,{\cal P}''_C,{\cal P}''_R ),
\end{equation}
where
\begin{equation}
  \mbox{minmod}(a,b,c)=\left\{
  \begin{array}{ccl}
     0 &\mbox{ if}& ab\le0  \mbox{ or}\\ 
       & &  bc\le 0\\
     min(a,b,c) &\mbox{ if}& a,b,c>0\\
     max(a,b,c) &\mbox{ if}& a,b,c<0
  \end{array} \right.
\end{equation}
The left and right  states of each cell interface that are found via this second reconstruction 
are then used to compute HLL-fluxes ${\cal F}_{m+\frac{1}{2},n+\frac{1}{2}}$ of equation (\ref{step2-int}). 
The resulting scheme is second order accuracy in time and third order 
accuracy in space.

\begin{figure*}
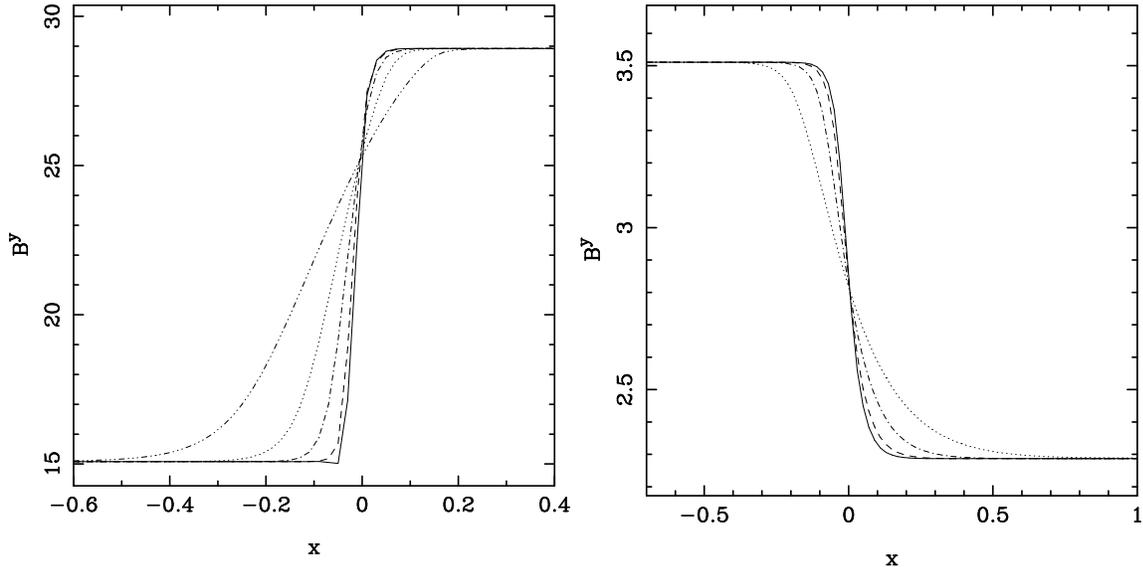

\includegraphics[angle=-90,width=75mm]{figures/sfs-var.eps}
\includegraphics[angle=-90,width=75mm]{figures/sss-var.eps}
\caption{Dependence of shock structure on resistivity. 
{\it Left panel:} Stationary fast shock for $\eta=0.01$ (solid line),
$\eta=0.03$ (dashed),$\eta=0.09$ (dash-dotted), 
$\eta=0.18$ (dotted), $\eta=0.36$ (dash-triple-dotted);
{\it Right panel:} Stationary fast shock for $\eta=0.01$ (solid line),
$\eta=0.02$ (dashed),$\eta=0.04$ (dash-dotted),
$\eta=0.08$ (dotted).
}
\label{var}
\end{figure*}

\begin{figure*}
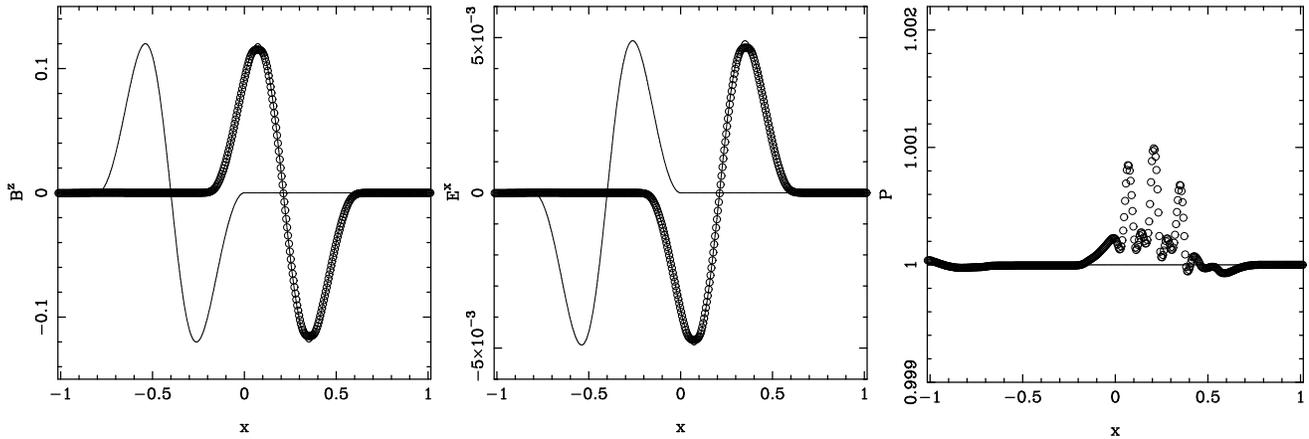

\includegraphics[angle=-90,width=57mm]{figures/aw-b3.eps}
\includegraphics[angle=-90,width=57mm]{figures/aw-e.eps}
\includegraphics[angle=-90,width=57mm]{figures/aw-p.eps}
\caption{ Alfv\'en wave. 
}
\label{aw}
\end{figure*}

\section{Test simulations}
\label{tests}

In these test simulations we use the polytropic equation of 
state 
\begin{equation}
  w = \rho +\frac{\Gamma}{\Gamma-1}p
\end{equation}
with the ratio of specific heats $\Gamma=4/3$.

\begin{figure}
\includegraphics[angle=-90,width=75mm]{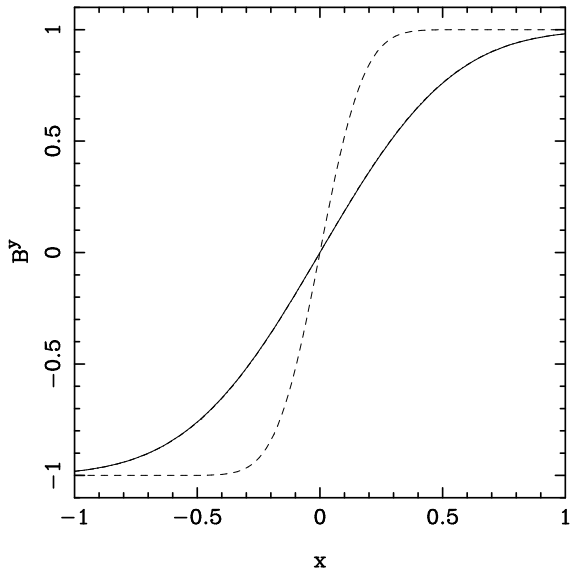}
\caption{ Self-similar current sheet. The dashed and the dash-dotted lines show the 
exact solution at $t=1$ and $t=9$ respectively. The solid line shows the numerical 
solution at $t=9$; it is indistinguishable from the exact solution on this plot.  
}
\label{errf}
\end{figure}

\subsection{One-dimensional test problems}

\subsubsection{Stationary Fast Shock}  

To set up this test we solved the ideal relativistic MHD shock equations describing 
stationary shocks. The selected particular solution is 
\vskip 0.3cm
\noindent
{\it Left state:}  \\$\bB=(5.0,15.08,0.0)$, $\gamma\bv=(4.925,0.0,0.0)$, $\rho=1.0$,\\ 
$p=10.0$, $q=0$, $\Phi=0$, $\Psi=0$;\\
\vskip 0.1cm
\noindent
{\it Right state:}  \\$\bB=(5.0,28.92,0.0)$, $\gamma\bv=(0.6209,0.1009,0.0)$, $\rho=7.930$,\\ 
$p=274.1$; $q=0$, $\Phi=0$, $\Psi=0$.
\vskip 0.3cm
    
\noindent
The electric field is found via the ideal equation 

\begin{equation}
  -\bE=-\vpr{v}{B}. 
\label{E-ideal}
\end{equation}
The computational grid is uniform and has 100 cells in $[-1,+1]$ and the initial solution is 
set as a discontinuity at $x=0$. The resistivity is $\eta=0.01$.
Figure~\ref{sfs} shows the numerical solution at $t=3.0$ by when the secondary waves created 
during the development of the dissipative shock structure have left the grid.    
One can see that the shock jump is captured very well.  The fact that there are only 3 grid
points in the shock structure tells that the shock is unresolved and suggests that the 
shock structure might be dominated by numerical dissipation. This is confirmed by the 
simulations with higher resistivity (see fig.\ref{var}).

\subsubsection{Stationary Slow Shock}
                                                                                                       
To set up this test we also solved the ideal relativistic MHD shock equations describing
stationary shocks. Now the selected particular solution is 
\vskip 0.3cm
                                                                                                       
\noindent
{\it Left state:}  \\$\bB=(5.0,3.511,0.0)$, $\gamma\bv=(0.6082,0.0,0.0)$, $\rho=1.0$,\\ 
$p=10.0$,  $q=0$, $\Phi=0$, $\Psi=0$;\\
\vskip 0.1cm
\noindent
{\it Right state:}  \\$\bB=(5.0,2.287,0.0)$, $\gamma\bv=(0.4096,-0.2147,0.0)$, $\rho=1.485$, 
$p=17.03$; $q=0$, $\Phi=0$, $\Psi=0$.
\vskip 0.3cm

\noindent
The computational grid is uniform and has 100 cells in $[-1,+1]$ and the initial solutions is
a discontinuity at $x=0$. The resistivity is $\eta=0.01$.
Figure~\ref{sss} shows the numerical solution at $t=4.0$ by when the secondary waves created 
during the development of the dissipative shock structure have left the grid.    
Again, the shock jump is captured very well but now there are more then 10 grid
point in the shock structure. This suggest that shock may be resolved. However, the rather 
small increase in the shock width between the cases with $\eta=0.005$ and $\eta=0.01$ 
shows that numerical dissipation is still important for $\eta=0.01$ and only for higher 
resistivity the shock structure becomes fully resolved (see fig.\ref{var}).  
                                                                                                   
\subsubsection{Alfv\'en wave}

To set up this test we utilized the analytical solution for ideal MHD Alfv\'en waves 
obtained in Komissarov\shortcite{Kom97}. In this test $\rho=1.0$, $p=1.0$, $B^x=1.0$, 
and the Alfv\'en speed $c_a=0.4079$. 
Initially the wave occupies the zone $x_0<x<x_1$, with $x_0=-0.8$, $x_1=0.0$. 
To the left of the wave $\bB=(1.0,0.1,0.0)$, $\gamma\bv=0$. In the wave 
the angle $\theta$ between the tangential component of magnetic field and the y-axis 
varies as 

$$
 \theta = 2\pi(3\xi^2-2\xi^3), \qquad \xi=(x-x_0)/(x_1-x_0),
$$
that gives vanishing first derivatives at $x_{0,1}$. The initial electric field is 
computed via eq.(\ref{E-ideal}) and the electric charge density via eq.(\ref{GaussE}).  
The computational grid is uniform and has 400 cells in $[-1,+1]$. 
The resistivity is set to a relatively small value, $\eta=0.003$, in order to get closer
to the ideal case. The simulations are continued up to $t =1.5$ and then compared with 
the exact solution of ideal MHD at the same time (fig.\ref{aw}). 
One can see that the agreement is pretty good.    
The ideal solution keeps the wave profile totally invariant, however the numerical solution 
is a little distorted, mainly due to numerical dissipation (this is confirmed by studying 
the dependence on $\eta$). 

When the zero gradient boundary conditions (free-flow) are utilised in the simulations then 
both the fast and the slow waves do not get 
reflected of boundaries and cleanly pass through. However, the Alfv\'en waves exhibit noticeable 
reflection (in contract to the results with our ideal MHD code). We have not figured out yet 
as to how to avoid such a reflection.

\begin{figure*}
\includegraphics[angle=0,width=75mm]{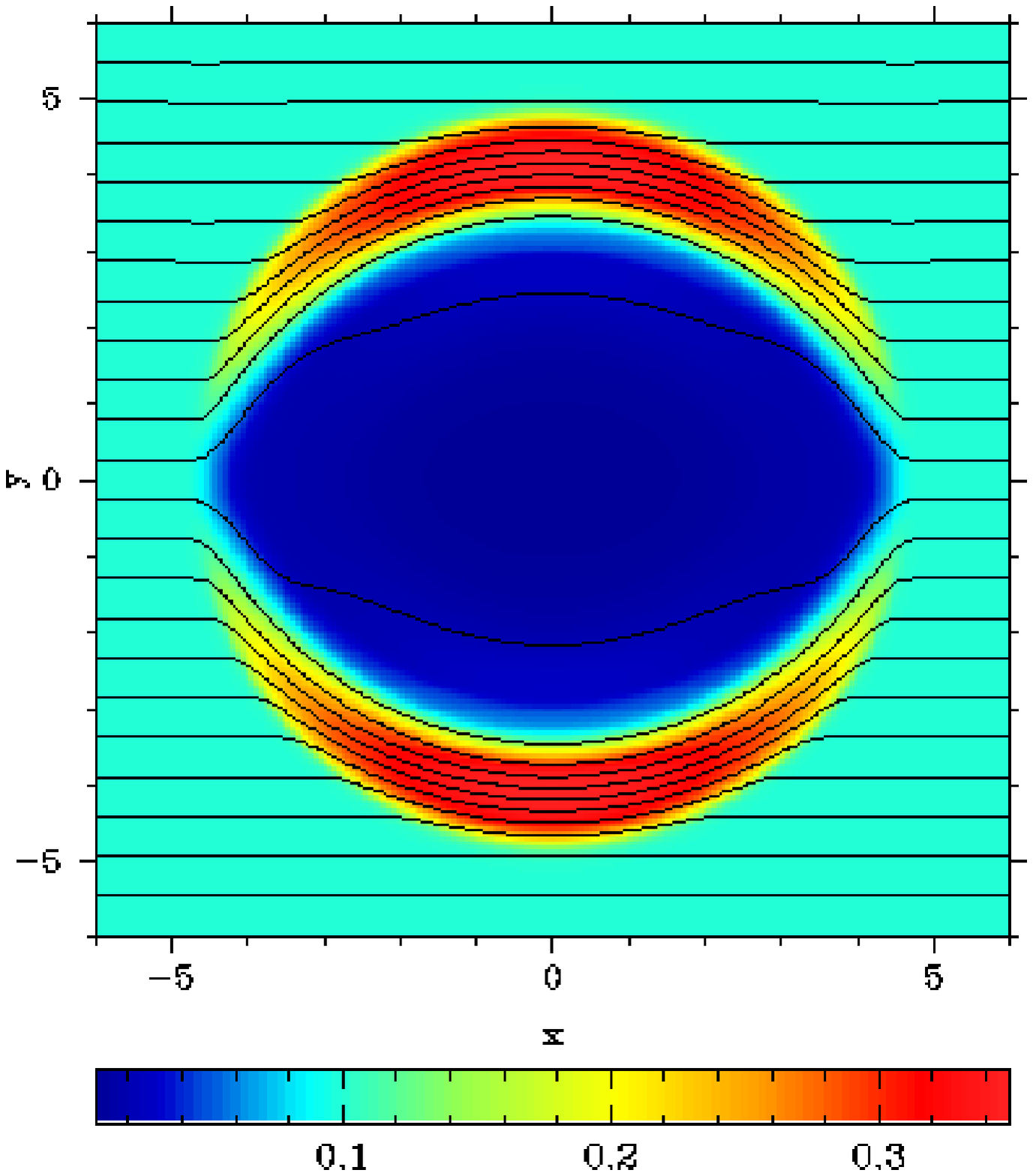}
\includegraphics[angle=0,width=75mm]{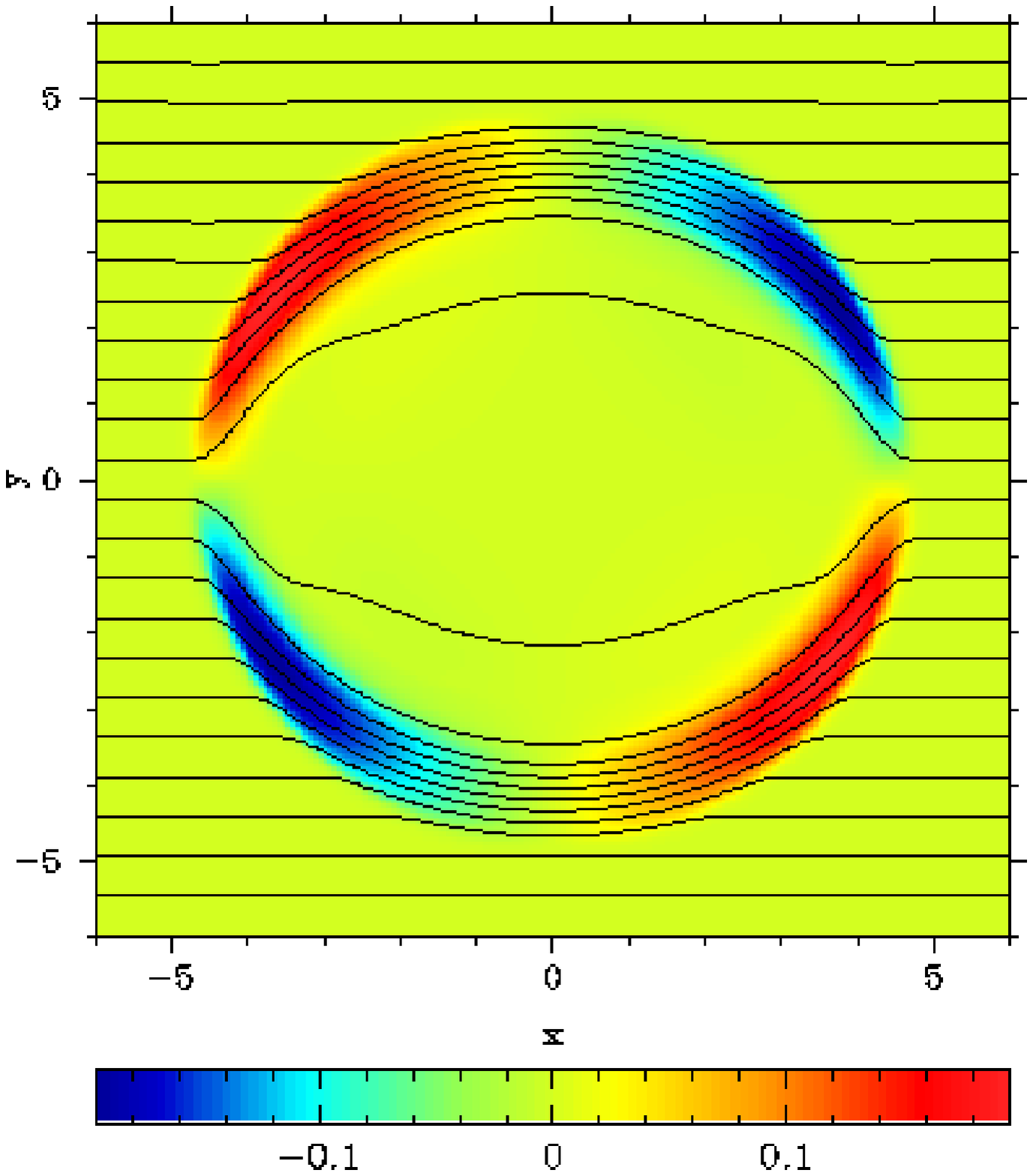}
\includegraphics[angle=0,width=75mm]{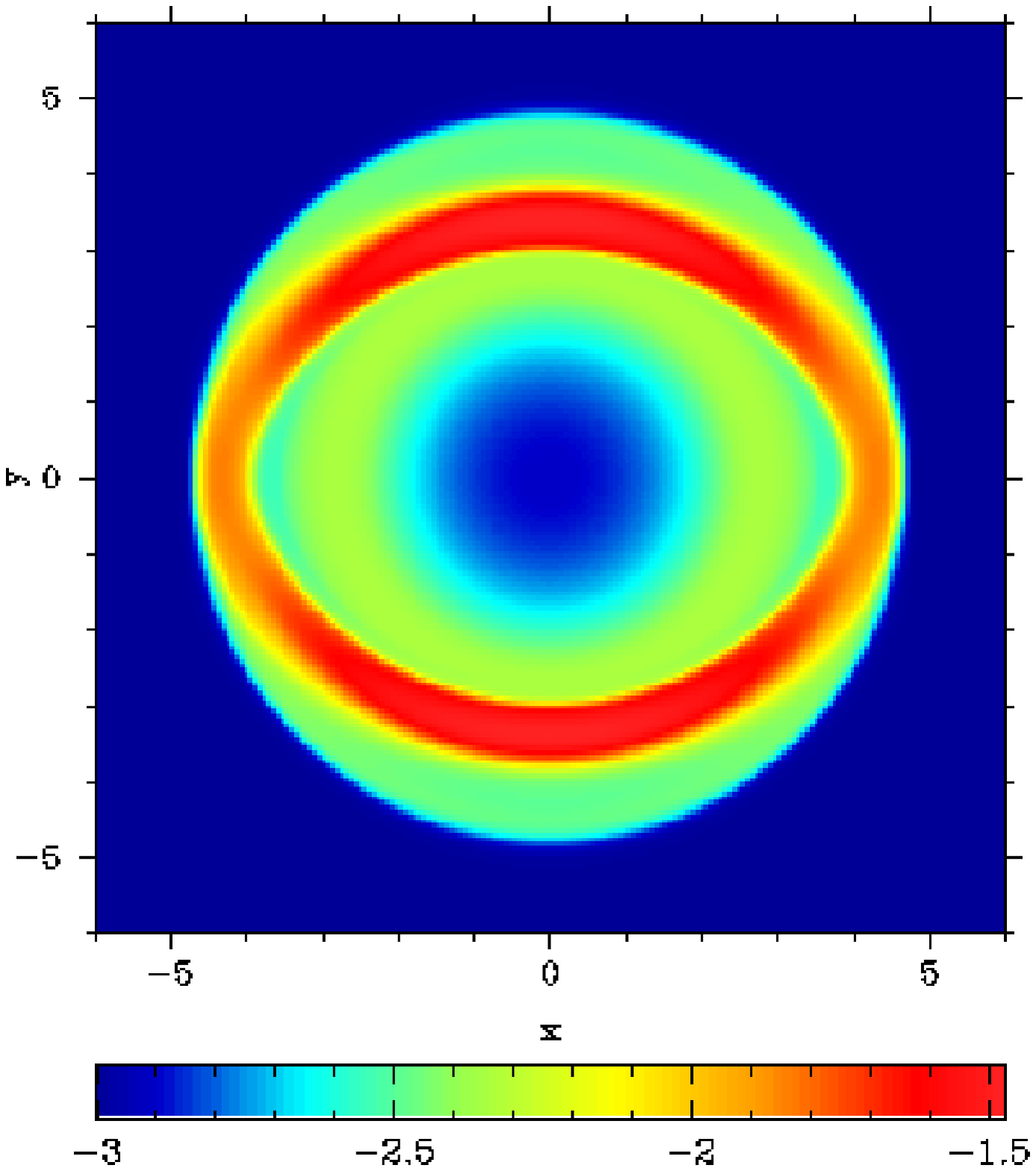}
\includegraphics[angle=0,width=75mm]{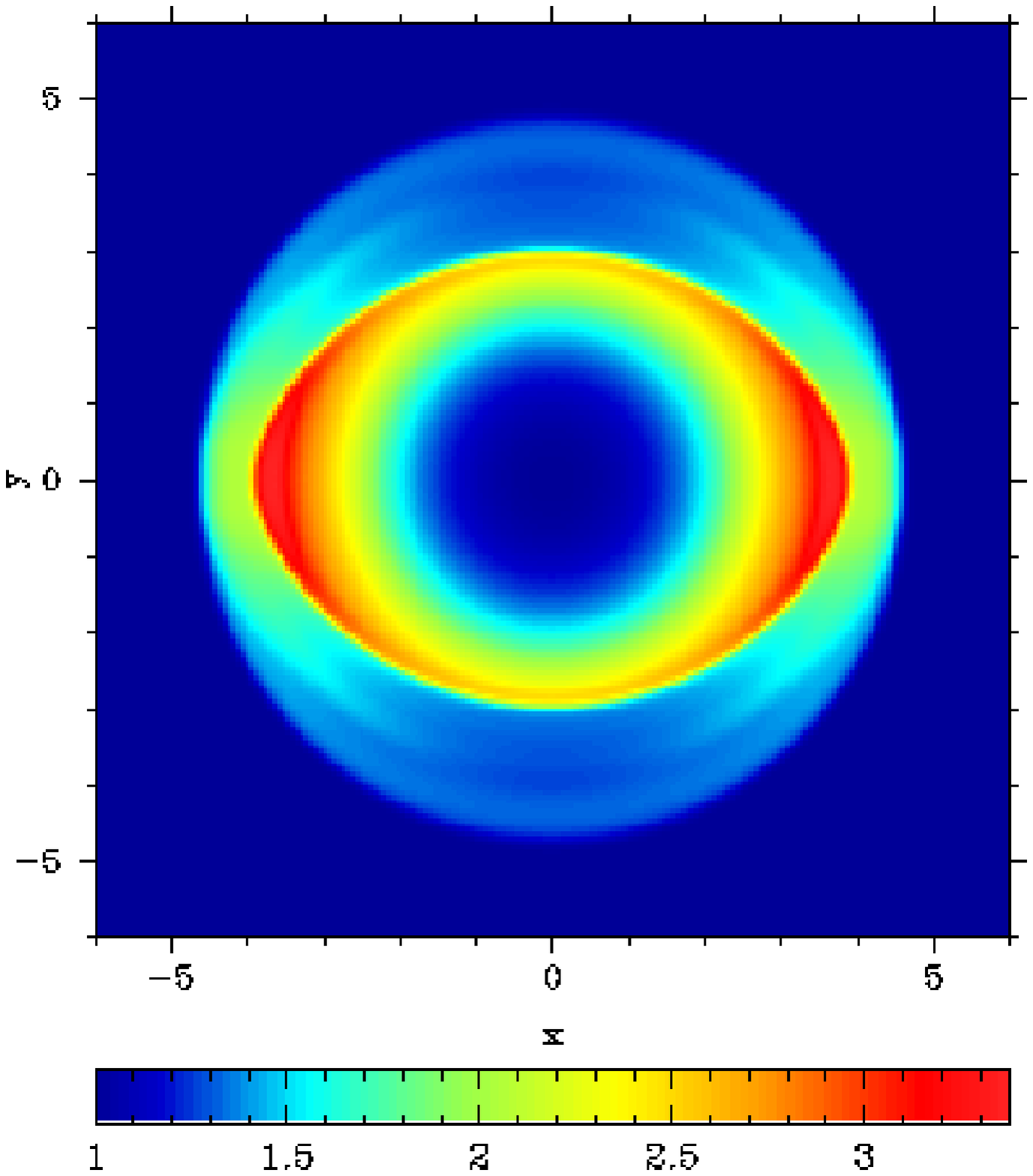}
\caption{ Strong cylindrical explosion. 
{\it Top left panel:} $B^x$ and magnetic field lines; 
{\it Top right panel:} $B^y$ and magnetic field lines;
{\it Bottom left panel:} $log_{10}p$, gas pressure;
{\it Bottom right panel:} Lorentz factor. 
}
\label{bw2d}
\end{figure*}

\begin{figure*}
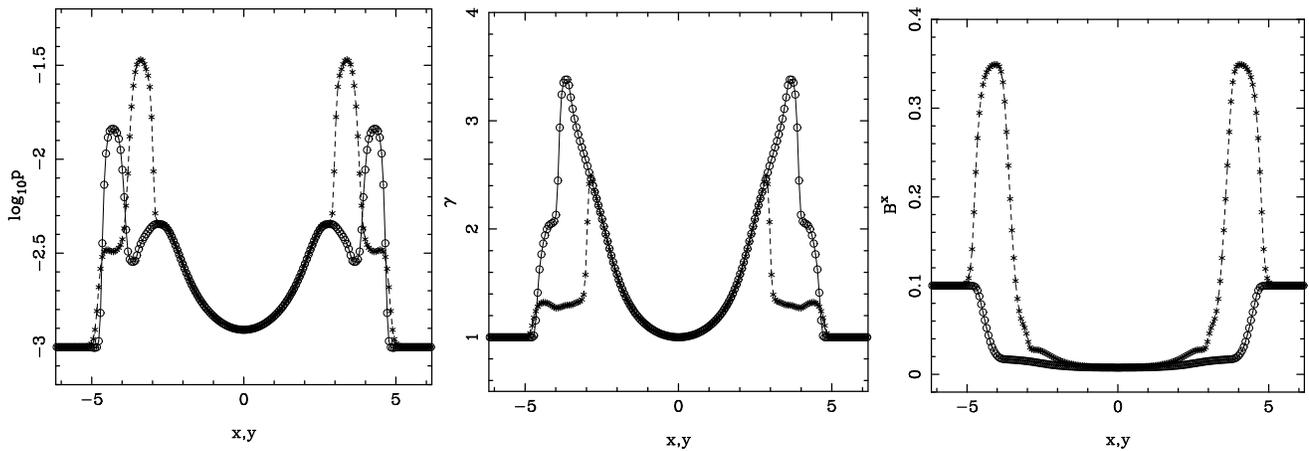

\includegraphics[angle=-90,width=57mm]{figures/bw2-cut1.eps}
\includegraphics[angle=-90,width=57mm]{figures/bw2-cut2.eps}
\includegraphics[angle=-90,width=57mm]{figures/bw2-cut3.eps}
\caption{Strong cylindrical explosion. This plot show slices along 
$x=0$ (dashed line and stars) and $y=0$ (solid lines and circles) for 
gas pressure (left panel), Lorentz factor (middle panel), and $B^x$ (right panel).
}
\label{bw2d-cut}
\end{figure*}

\subsubsection{Self-similar current sheet}

Assume that $\bB=(0.0,B(x,t),0.0)$, the magnetic pressure 
is much smaller than the gas pressure everywhere, and $B(x,0)$ changes sign within a thin 
current layer of width $\Delta l$. Provided the initial solution is in equilibrium, 
$p=$const, the evolution is a slow diffusive expansion of the layer caused by the resistivity and 
described by the archetypal diffusion equation

$$
  \Pd{t}B -\eta\Pd{x}^2B =0.
$$
As the width of the layer becomes much larger than $\Delta l$ the expansion becomes self-similar 
\begin{equation}
   B(x,t)=B_0\, \mbox{erf}\left(\frac{1}{2\sqrt{\eta\xi}}\right), \qquad \xi=t/x^2, 
\label{erf}
\end{equation}
where erf is the error function, and this analytic results can be used to test the resitive 
part of the code. In the test problem that is presented here the initial solution has uniform 
distribution of $P=50.0$, $\rho=1.0$, $\bE=0$, and $\gamma\bv=0$ and the initial magnetic field 
is given by eq.(\ref{erf}) for $B_0=1.0$, $t=1$, and $\eta=0.01$.  
The computational grid is uniform and has 200 cells in $[-1.5,+1.5]$. 
The numerical simulations are continued up to $t=8$ and then the numerical solution is 
compared with the solution (\ref{erf}) for $t=9$. The results are shown in figure~\ref{errf} -- 
one cannot see the difference between the solutions.

\begin{figure*}
\includegraphics[angle=0,width=75mm]{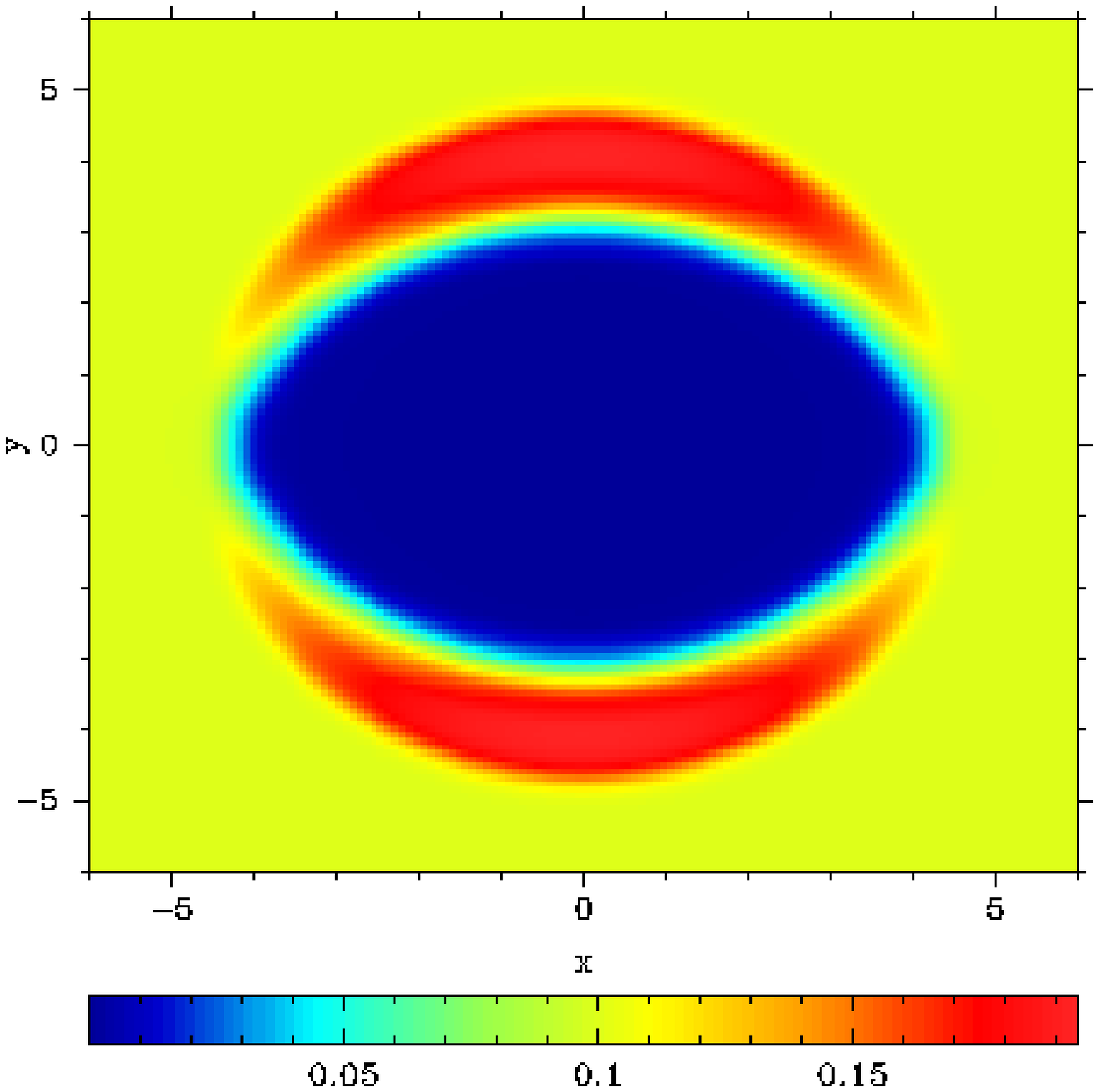}
\includegraphics[angle=0,width=75mm]{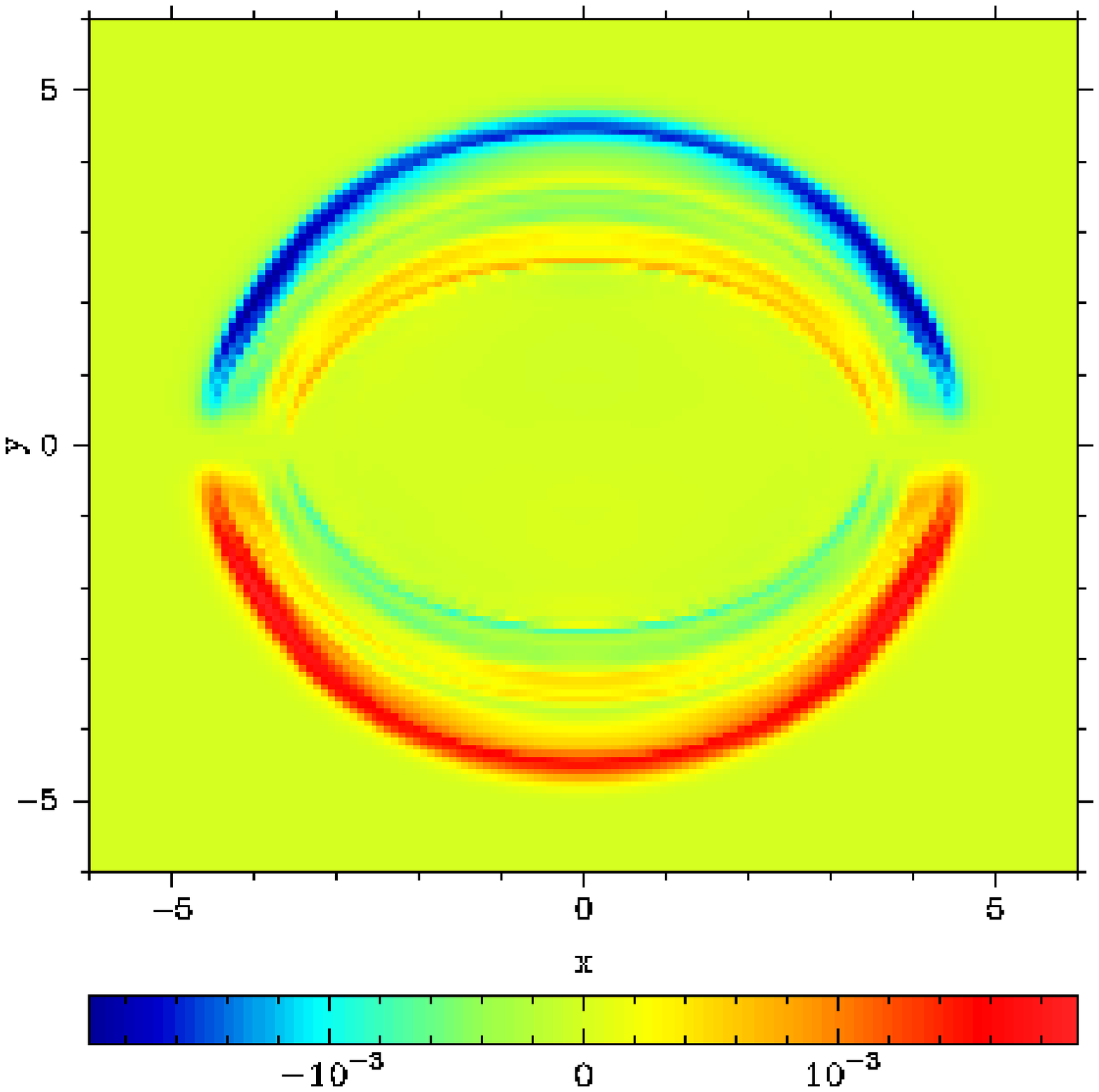}
\includegraphics[angle=0,width=75mm]{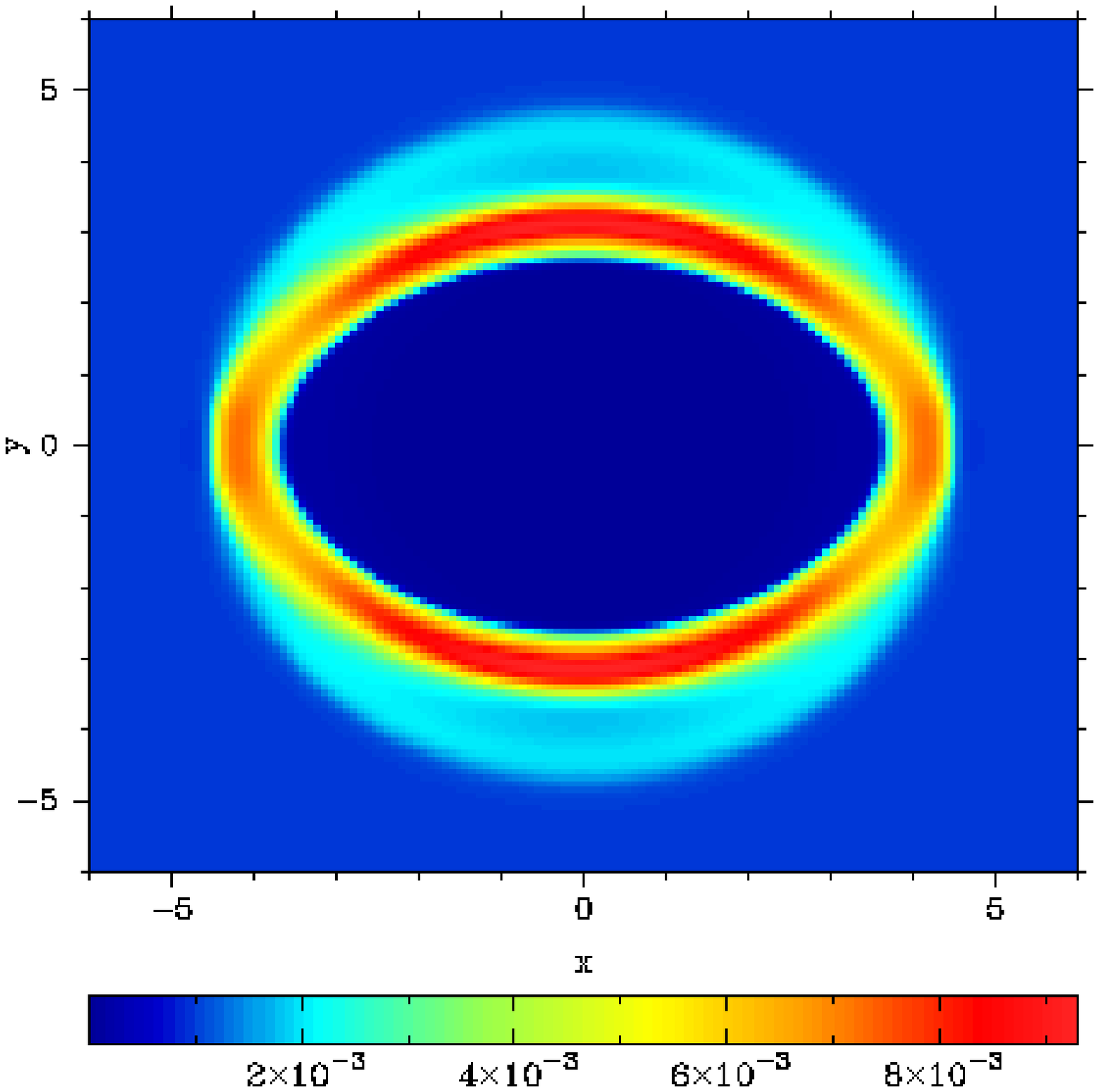}
\includegraphics[angle=0,width=75mm]{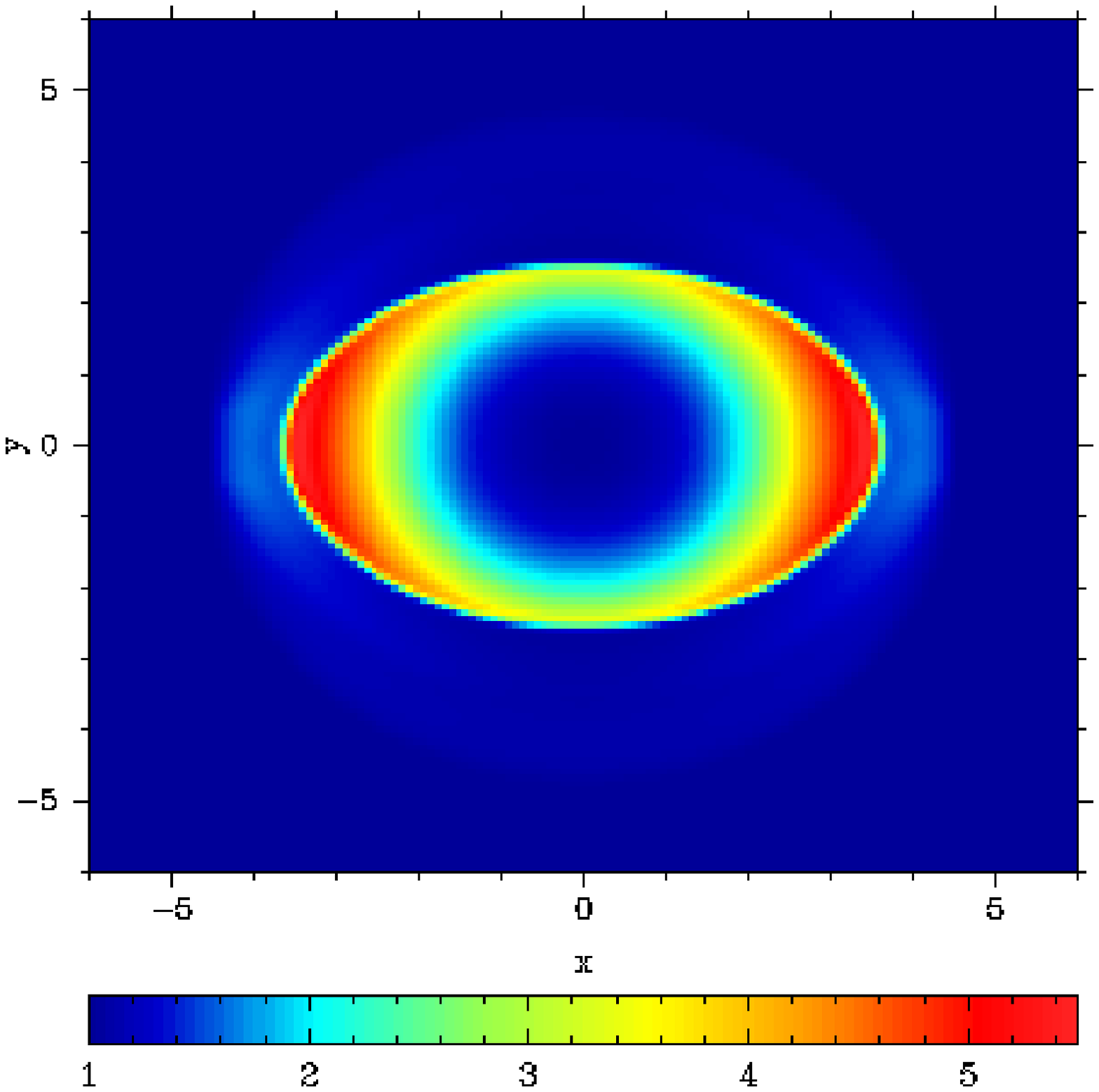}
\caption{ Strong spherical explosion.
{\it Top left panel:} $B^x$;
{\it Top right panel:} $q$, electric charge density; 
{\it Bottom left panel:} $p$, gas pressure;
{\it Bottom right panel:} Lorentz factor.
}
\label{bw3d}
\end{figure*}

\begin{figure*}
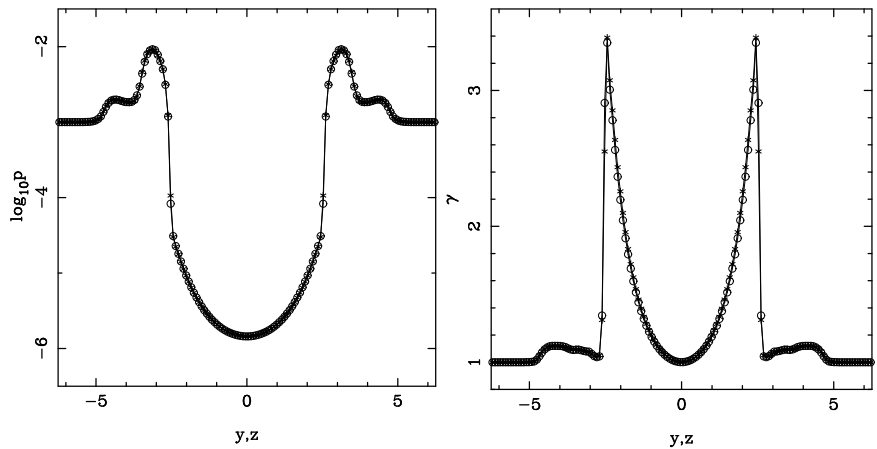

\includegraphics[angle=-90,width=57mm]{figures/bw3d-cut-p.eps}
\includegraphics[angle=-90,width=57mm]{figures/bw3d-cut-g.eps}
\caption{ Strong spherical explosion. In both panels stars show the 
solution along the 
{\it Left panel:}  $log_{10}p$, gas pressure. 
{\it Right panel:} $q$, electric charge density.
}
\label{bw3d-cut}
\end{figure*}

\subsection{Multi-dimensional tests}

All the one-dimensional problems, that are described above, have been used to test both 
the 2D and 3D versions of the code via application in all two/three directions of the Cartesian 
grid.  The results are almost identical to that of 1D tests. In addition we considered 
several generically multi-dimensional problems.

\subsubsection{Strong cylindrical explosion}

Strong symmetric explosions are useful standard tests for MHD codes even if there are 
no exact analytic solutions to work with. This is because the generated shocks make all
possible angles to the grid and to the magnetic field thus allowing to detect well 
hidden bugs and to reveal potential weaknesses.  In this problem the Cartesian 
computational domain is $(-6.0,+6.0)\!\times\!(-6.0,+6.0)$ with 200 equidistant grid points in 
each direction. The initial explosion zone is a cylinder of radius $r=1$ centered onto the 
origin. Its pressure and density are set to $p=1$ and $\rho=0.01$ for $r<0.8$ and 
exponentially decrease for $0.8<r<1.0$. The ambient gas has $p=\rho=0.001$. The initial
magnetic field is uniform, $\bB=(0.1,0.0,0.0)$. Figure~\ref{bw2d} shows the 2D solution at 
$t=4$ for $\eta=0.018$ and $\eta_d=1/\kappa=0.18$. 
It exhibits the same features as the ideal MHD solution of a 
similar test problem \cite{Kom99} which is expected given the low value of $\eta$ and 
shows nothing that could be suspected as artifacts.  
For more detailed future comparisons with other codes figure~\ref{bw2d-cut} shows 
slices of the solution along $x=0$ and $y=0$.          

The same problem has been used to test the 3D code with identical results.

\subsubsection{Strong spherical explosion}

Finally, we tested our 3D code on the problem of spherical explosion. 
All parameters of the explosion are the same as in the cylindrical case 
with exception of the explosion zone - now this is a sphere of unit radius. The 
computational domain is $(-6.0,+6.0)\!\times\!(-6.0,+6.0)\!\times\!(-6.0,+6.0)$ 
with 140 equidistant grid points in each direction.  
Figures \ref{bw3d} and \ref{bw3d-cut} show the numerical solution for 
$\eta=0.0257$ and $\eta_d=1/\kappa=0.257$ at $t=4.0$.      
The general structure of the solution is similar to that of the cylindrical 
case but with much stronger central rarefaction. One qualitatively new 
feature is the non-vanishing electric charge density (top right panel of fig.\ref{bw3d}).
Given the axial symmetry of the problem one expects the solutions to be the same 
in the planes $z=0$ and $y=0$. Figure~\ref{bw3d-cut} shows that this is indeed the case.

\section{Conclusions}
\label{conclusions}

We have constructed a multidimensional upwind scheme for resistive 
relativistic magnetohydrodynamics. At the moment only the case of scalar 
resistivity has been implemented and more work has to be done to incorporate 
the case of tensor resistivity. The results of test simulations show that the 
scheme is robust in the regime of small to moderate magnetization, which can be 
described  by the ratio of the electromagnetic energy density to the total mass-energy 
density of matter. The regime of high magnetization is still problematic as the 
the truncation errors for the energy-momentum of matter become large often making 
impossible to convert the conserved quantities into the primitive ones. This is 
a well know problem of all conservative schemes for relativistic MHD. Apart from this 
drawback the scheme can handle equally well both resistive current sheets and shock waves 
and thus can be a useful tool for studying phenomena of relativistic astrophysics that involve
both colliding supersonic flows and magnetic reconnection.

\section*{Acknowledgments}
This research is funded by PPARC under the rolling grant
``Theoretical Astrophysics in Leeds''. The author thanks Matthew Boham 
for his help in testing the 3D code and Maxim Barkov who is currently working 
on its parallelization.



\begin{thebibliography}{}
\bibitem[\protect\citename{Anderson et~al. }2006]{AHLN06}
 Anderson M., Hirschmann E.W., Liebling S.L., Neilsen D., 2006,CQGra,23,6503
\bibitem[\protect\citename{{Anninos} et~al. }2006]{Ann06}
 Anninos P., Fragile P.~C., Salmonson J.~D., 2005, ApJ, 635, 723
\bibitem[\protect\citename{Ant\'on et~al. }2006]{Ant06}
 Ant\'on L., Zanotti O., Miralles J.~A., Mart\'{i} J. M., Ib\'a\~nez
 J.~M., Font J.~A., Pons J.~A.,2006,ApJ,637,296
\bibitem[\protect\citename{Blackman \& Field }1993]{BF93}
 Blackman E.G., Field G.B., 1993, Phys.Rev.Lett.,71,3481 
\bibitem[\protect\citename{De Villiers \& Hawley }2003]{DH03}
 De Villiers J.-P., Hawley J.F., 2003,ApJ,589,458.
\bibitem[\protect\citename{{Del Zanna} et~al. }2003]{Del03}
 Del Zanna L., Bucciantini N., Londrillo P., 2003,A\&A,400,397
\bibitem[\protect\citename{Del Zanna et al. }2007]{Del07}
 Del Zanna L., Zanotti O., Bucciantini N., Londrillo P., 
 2007, submitted to A\&A (arXiv:0704.3206)
\bibitem[\protect\citename{Duez et~al. }2005]{Due05}
 Duez M.D., Liu Y.T., Shapiro S.L., Stephens B.C., 2005,Phys.Rev.D,72,024028
\bibitem[\protect\citename{Gammie et~al. }2003]{GMT03}
 Gammie C.F., McKinney J.C., Toth G.,2003,ApJ,589,444.
\bibitem[\protect\citename{Godunov }1959]{God}
 Godunov S.K.,1959,Mat.Sb.,47,357.
\bibitem[\protect\citename{Giacomazzo \& Rezzolla }2007]{GR07}
 Giacomazzo B., Rezzolla L., 2007, Class.Quant.Grav.,24(12),S235
\bibitem[\protect\citename{Harten et~al. }1983]{HLL}
 Harten A., Lax P.D., van Leer B., 1983, SIAM Rev,25,35.
\bibitem[\protect\citename{e.g. Jackson }1979]{JAC}
 Jackson J.D., 1979, {\it Classical Electrodynamics}, John Wiley \& Sons, New York
\bibitem[\protect\citename{Koide et~al. }1999]{KSK99}
 Koide S., Shibata K., Kudoh T., 1999, ApJ, 522, 727
\bibitem[\protect\citename{Koldoba et~al. }2002]{KKU}
 Koldoba A.V., Kuznetsov O.A., Ustyugova G.V., 2002, MNRAS, 333, 932.
\bibitem[\protect\citename{Komissarov }1997]{Kom97} 
 Komissarov S.S., 1997,Phys.Lett.A,232,435
\bibitem[\protect\citename{Komissarov }1999]{Kom99} 
 Komissarov S.S., 1999,MNRAS,303,343
   Koide S., Shibata K., and Kudoh T., 1999, Ap.J., {\bf 522}, 727.
\bibitem[\protect\citename{Komissarov }2001]{Kom01}
   Komissarov S.S., 2001, in ``Godunov Methods: Theory and Applications'',
   ed. E.F.Toro, Kluwer, New York, p.519.
\bibitem[\protect\citename{Komissarov }2004]{Kom04}
 Komissarov S.S., 2004,MNRAS,350,1431.
\bibitem[\protect\citename{Lyubarsky }2005]{Lyub05}
 Lyubarsky Y.E., 2005, MNRAS,358,113.
\bibitem[\protect\citename{Lyutikov \& Uzdensky }2003]{LU03}
 Lyutikov M., Uzdensky D., 2003,ApJ,589,893.
\bibitem[\protect\citename{McKinney }2006]{M06}
 McKinney J.C., 2006,MNRAS,367,1797.
\bibitem[\protect\citename{Mignone \& Bode }2006]{MB06}
 Mignone A., Bodo G., 2006,MNRAS,368,1040.
\bibitem[\protect\citename{Mignone et al. }2007]{Mig07}
 Mignone A., Bodo G., Massaglia S., Matsakos T., Tesileanu O., Zanni C., Ferrari A.,
 2007,ApJS,70(1),228.
\bibitem[\protect\citename{Mizuno et al. }2007]{Miz06}
 Mizuno Y., Nishikawa K.-I., Koide S., Hardee P., Fishman G.J.,
 2006, submitted to ApJS (astro-ph/0609004).
\bibitem[\protect\citename{Munz et al. }1999]{Munz99} 
 Munz C.-D.,Omnes P., Schneider R., Sonnendruker E., Vob U., 
 1999,JCP,161,484.
\bibitem[\protect\citename{Neilsen et al. }2006]{NHM06}
 Neilsen D., Hirschmann E.W., Millward R.S., 2006,Class.Quant.Grav.,23(16),S505.
\bibitem[\protect\citename{Noble et al. }2006]{N06}
 Noble S.C., Gammie C.F., McKinney, J.C., Del Zanna L., 2006,ApJ,641,626.
\bibitem[\protect\citename{{Shibata \& Sekuguchi }}2005]{SS05}
 Shibata M., Sekuguchi Y. I., 2005,Phys.Rev.D.,72,044014
\bibitem[\protect\citename{Strang} 1968]{Str} 
 Strang G., 1968,SIAM J.Num.Anal.,5,506
\bibitem[\protect\citename{Watanabe \& Yokoyama} 2006]{WY06}
 Watanabe N., Yokoyama T.,2006,ApJ,647,L123.
 
\end{thebibliography}
\end{document}